\documentstyle[12pt]{article}
\newcommand{\Tr}[1]{\,{\rm Tr}\,#1\,}

\begin{document}
\title{
\begin{flushright}
{\small SMI-9-93 }
\end{flushright}
\vspace{0.5cm}
Manifestly Gauge Invariant Models of Chiral Lattice Fermions}
\author{ A.A.Slavnov
\thanks{E-mail:$~~$ slavnov@qft.mian.su}
\\ Steklov Mathematical Institute, Russian Academy of Sciences,\\
Vavilov st.42, GSP-1,117966, Moscow, Russia }
\maketitle
\begin{abstract}
A manifestly gauge invariant lattice action for nonanomalous chiral models
is proposed which leads in the continuum limit to the theory free of
doublers.

\end{abstract}

\section   {Introduction}
In this paper I present a gauge invariant lattice formulation of the
anomaly free chiral models like the $SO(10)$ or the standard model. It will
be proven in the framework of perturbation theory that all the fermion
doublers decouple in the continuum limit and the physical content of the
the theory is the usual one of the standard model.

The present model is a gauge invariant extension of the model proposed in
our paper \cite{SF}, in which the gauge invariance was broken for a finite
lattice spacing and restored in the continuum limit.

To provide the doublers with big masses we use the Yukawa--Wilson
interaction of fermions with Higgs mesons. This possibility was widely
discussed last years mainly in the framework of the Smit-Swift model
\cite{Kars}, \cite{Smit1}, \cite{Swift}, \cite{Smit2}. (For detailed
references  and review of other approaches see \cite{Smit3}, \cite{Golt},
\cite{Pet}).

However the models used so far did not solve the problem.To give the doublers
big masses one needs a large Yukawa coupling. It prevents using the
perturbation theory as a big Yukawa coupling produces strong effects on
the Higgs mesons interaction. On the other hand the strong coupling
analysis leads to the conclusion that the light fermions in these models
are noninteracting \cite{GPS}, \cite{BS}, \cite{GP}.

In this paper I shall show that if one introduces the additional gauge
invariant interaction with the Pauli-Villars (PV) fields, and higher
covariant derivative kinetic terms for the Higgs and Yang-Mills fields,the
effective Yukawa-Wilson coupling becomes weak and the the model can be
treated perturbatively.

\section {The Model}
We consider the $SO(10)$ model described by the Lagrangian:
\begin{equation}
L=-1/4(F_{\mu\nu}^{ij})^2+i\sum_l\bar{\psi}_+^l(\hat{\partial}-ig\hat{A}^{ij}\sigma_{ij})\psi_+^l+L_H
\label{d1} \end{equation}
Here $\sigma_{ij}$ are the $SO(10)$ generators:
$\sigma_{ij}=1/2[\Gamma_i,\Gamma_j]$, where $\Gamma_i$ are Hermitian
$32\times32$ matrices which satisfy the Clifford algebra:
$[\Gamma_i,\Gamma_j]=2\delta_{ij}$. The chiral $SO(10)$ spinors
$\psi_\pm=1/2(1\pm\Gamma_{11})\psi$, where $\Gamma_{11}=\Gamma_1\Gamma_2
\ldots\Gamma_{10}$, describe the 16-dimensional irreducible representation
of $SO(10)$ including quark and lepton fields. We assume also that the spinors
$\psi_+$ are left-handed $\psi_+=1/2(1-\gamma_5)\psi_+$. Index $l$ numerates
generations. $L_H$ includes interactions of Higgs mesons and will not be
written explicitely.

We start with the analysis of one loop fermion diagrams. Let us take as
the gauge invariant regularization of the Lagrangian (\ref{d1}) the
following lattice action
\begin{equation} I=I_{YM}+I_\psi+I_{YW}+I_H \label{d2} \end{equation}
Here $I_{YM}$ is the standard Wilson lattice action for the $SO(10)$ gauge
fields, $I_\psi$ describes the gauge invariant interaction of the gauge
fields with the original fermions $\psi_l$, and also with the fermionic,
$\psi_r$, and bosonic $\chi_r$, $\bar{\chi}_r$, PV fields
\begin{equation}
I_\psi= \sum_{x,\mu ,l} \left[ -\frac 1{2ia} \bar{\psi}_l^+ (x) \gamma^\mu
U_\mu (x) \psi_l^+ (x+a_\mu) \right]
\label{d3}
\end{equation}
$$
+\sum_{x,\mu,r} \left[ -\frac 1{2ia} \bar{\psi}_r (x) \gamma^\mu
U_\mu (x) \psi_r (x+a_\mu) - \frac {M_r}2 \bar{\psi}_r (x) C_D C
\Gamma_{11} \bar{\psi}_r^T (x) \right]
$$

$$
+\sum_{x,\mu,r} \left[ -\frac 1{2ia} \bar{\chi}_r (x) \gamma^\mu
\Gamma_{11} U_\mu (x) \chi_r (x+a_\mu) - \frac 1{2ia} \bar{\tilde{\chi}}_r (x)
\Gamma_{11} U_\mu (x) \tilde{\chi}_r (x+a_\mu)\right]
$$

$$
+\frac {M_r}2 \left[ \bar{\chi}_r (x) C_D C
 \bar{\chi}_r^T (x) + \bar{\tilde{\chi}}_r (x) C_D C
\bar{\tilde{\chi}}_r^T (x) \right] +h.c.
$$
Here $U_\mu(x)$ is the usual lattice gauge field
$U_\mu=\exp\left\{ig\sigma_{ij}A_\mu^{ij}\right\}$. The matrix $C$ is a
conjugation matrix for the $SO(10)$ group: $C\sigma = -\sigma^{T}C$, the
matrix $C_D$ is the usual charge conjugation matrix. The terms
proportional to $M_r$ are the gauge invariant Majorana mass terms for the
PV fields. The number of PV fields with the mass $M_r$ will be denoted by
$C_r$.Contributions of bosonic and fermionic PV fields to spinorial loops
differ by sign.

$I_{YW}$ is the Yukawa--Wilson action which produces via spontaneous
symmetry breaking the Wilson like mass terms lifting the values of the
doublers masses:
\begin{equation}
I_{YW}= - \frac N{2} \sum_{l,x,\mu} \Bigl[
\bar{\psi}_l^+ (x) \varphi^i (x)\Gamma^i
U_\mu (x) C_D C\bar{\psi}_l^{+T} (x+a_\mu)
\label{d4}
\end{equation}
$$
+\bar{\psi}_l^+ (x+a_\mu) \Gamma^i C_D C\varphi^i (x+a_\mu) U_\mu^T (x)
 \bar{\psi}_l^{+T} (x) -2\bar{\psi}_l^+ (x) \Gamma^i C_D C\varphi^i
(x) \bar{\psi}_l^{+T} (x) \Bigr] $$

$$
- \frac N{2} \sum_{r,x,\mu} \Bigl[
\bar{\psi}_r (x) \varphi^i (x)\Gamma^i
U_\mu (x) C_D C\bar{\psi}_r^T (x+a_\mu)
$$

$$
+\bar{\psi}_r (x+a_\mu) \Gamma^i C_D C\varphi^i (x+a_\mu) U_\mu^T (x)
\bar{\psi}_l^T (x) -2\bar{\psi}_r (x) \Gamma^i C_D C\varphi^i (x)
\bar{\psi}_r^T (x) \Bigr]
$$

$$
- \frac N{2} \sum_{r,x,\mu} \Bigl[
\bar{\tilde{\chi}}_r (x) \varphi^i (x)\Gamma^i
U_\mu (x) C_D C\bar{\chi}_r^T (x+a_\mu) +\bar{\tilde{\chi}}_r (x+a_\mu)
\varphi^i (x+a_\mu) \Gamma^i C_D C U_\mu^T (x)  \bar{\chi}_r^T (x)
$$

$$
 -2\bar{\tilde{\chi}}_r (x)
 \Gamma^i C_D C\varphi^i (x) \bar{\chi}_r^T (x) \Bigr]+h.c.
$$
Here $\varphi^i$ is the Higgs field realising the 10-dimensional
representation of $SO(10)$. $N$ is a large dimensionless parameter which
becomes infinite in the continuum limit. We shall put $a^{-1} = \lambda
N$, where $\lambda$ is a fixed mass scale.

Finally $I_H$ is the Higgs action including the gauge invariant kinetic
term and the Higgs potential which we do not write explicitely. We also
did not introduce  the scalar--fermion interaction producing finite
masses for physical leptons and quarks. It can be done in a standard way
and will not be discussed here.

A nonzero expectation value of the Higgs field $<\varphi_{10}>=v$ produces
the Wilson terms giving all the doublers the big masses
$\sim Nv$. The price we pay for that is the presence of the
large Yukawa coupling $N\sim \lambda a^{-1}$. Due to this coupling the heavy
particles may not decouple in the framework of perturbation theory and in
the limit $a\rightarrow 0$ new vertices including scalar and vector fields
may arise leading to breakdown of the weak coupling expansion.

We shall show that in our model due to the presence of the PV fields there
is a compensation between the contributions of bosonic and fermionic
fields leading to decoupling of the heavy particles and the absence of
additional vertices.

Let us consider the Higgs meson scattering amplitudes generated  by the
Yukawa interaction (\ref{d4}). For the diagram with L external lines we
have:
  \begin{equation}
\Pi_L \sim N^L \int\limits_{-\frac
{\pi}a}^{\frac {\pi}a}\, dp \Tr \{ \Bigl( \frac {1-\gamma_5}2 \Bigr)
\Bigl[ l\frac {(1+\Gamma_{11})}2 \ldots V^{i_L} S (p+Q_{L-1}) \ldots
V^{i_1} S (p)
\label{d5}
\end{equation}
$$
+ \sum_{r, \pm} \frac {(1\pm \Gamma_{11})}2 V^{i_L} S_r (p+Q_{L-1})
\ldots V^{i_1} S_r (p) \Bigr]
\}
$$

$Q_i=\sum_{n=1}^{i} k_n$  where $k_n$ are the external momenta. Here the
factor $l$ in the first term is due to the presence of $l$ generations of
original fields.

 $V^i$ is the interaction vertex
\begin{equation}
V^{i_t} =\Gamma^{i_t} C_D C \sum_{\mu} \left( \cos (p+Q_{L-1})_\mu a -1\right)
\label{d6}
\end{equation}
and $S_r$ are the propagators
\begin{equation}
S_{\bar{\psi}_+^l \psi_+^l}= \frac {\hat{s}}{s^2 +m^2},
\label{d7}
\end{equation}

\begin{equation}
S_{\bar{\psi}_r^+ \psi_r^+}= S_{\bar{\psi}_r^- \psi_r^-}=
S_{\bar{\chi}_r^+ \chi_r^+}= S_{\bar{\tilde{\chi}}_r^+ \tilde{\chi}_r^+}=
=-S_{\bar{\chi}_r^- \chi_r^-}= -S_{\bar{\tilde{\chi}}_r^- \tilde{\chi}_r^-}=
\frac {\hat{s}}{s^2 +m^2+M_r^2},
\label{d8}
\end{equation}

\begin{equation}
S_{\bar{\psi}_l^+ \bar{\psi}_l^+}= S_{\psi_l^+ \psi_l^+}= \frac
{\Gamma_{10} C C_D m}{s^2 + m^2+M_r^2 },
\label{d9}
\end{equation}

\begin{equation}
S_{\bar{\psi}_r^+ \bar{\psi}_r^+}= S_{\psi_r^+ \psi_r^+}=
S_{\bar{\psi}_r^- \bar{\psi}_r^-}= S_{\psi_r^- \psi_r^-}=
S_{\bar{\chi}_r^+ \tilde{\chi}_r^+}= S_{\bar{\chi}_r^- \tilde{\chi}_r^-}=
\frac {\Gamma_{10} C C_D m}{s^2 +\left( m^2+M_r^2\right) },
\label{d10}
\end{equation}

\begin{equation}
S_{\bar{\psi}_r^- \bar{\psi}_r^+}= S_{\psi_r^+ \psi_r^-}=
S_{\bar{\chi}_r^- \bar{\chi}_r^+}=
S_{\bar{\tilde{\chi}}_r^- \bar{\tilde{\chi}}_r^+}=
S_{\chi_r^+ \chi_r^-}= S_{\tilde{\chi}_r^+ \tilde{\chi}_r^-}=
\frac {M_r C_D C \Gamma_{11}}{s^2 +\left( m^2+M_r^2\right) },
\label{d11}
\end{equation}
Here
\begin{equation}
s_\mu= a^{-1} \sin (p_\mu a),
\label{12}
\end{equation}

\begin{equation}
m= Nv \sum_{\mu} (1-\cos (p_\mu a)).
\label{13}
\end{equation}

Note that the contribution of the original fields includes only positive
chirality projections $1/2(1+\Gamma_{11})$ whereas the PV fields have both
positive and negative chirality projections.

Let us show that in the limit $a\rightarrow 0$ the heavy particles
decouple. It is convinient to separate the integration domain in
(\ref{d5}) into two parts
\begin{equation}
V_{in} :~~\left| p\right| <\lambda N^{\gamma};~~~~V_{out} :  \left|
p\right| >\lambda N^{\gamma};~~~\gamma< \frac 13
\label{d14}
\end{equation}

In the domain $V_{in}$ $\mid ap\mid \ll 1$ and one can use the
expansion over $(ap)$. For example the diagrams including the propagators
(\ref{d7}),(\ref{d8}) look as follows
\begin{equation}
\Pi_{(in)}^L \sim N^{-L} \int_{V_{in}} \, \frac {\Tr [
\left( p+Q_{L-1}\right)^2  \left( \hat{p}+ \hat{Q}_{L-1}\right) \ldots p^2
\hat{p} \left( \frac {1+\gamma_5}2\right) ] \, d^4 p}{ [ \left(
p+Q_{L-1}\right)^2 +M_r^2 a^2 ] \ldots [ p^2+ M_r^2 a^2 ]
}\label{d15}
\end{equation}
One sees that $\Pi^L_{in} < N^{(4+L)\gamma -L}$ and $\Pi^L_{in}
\rightarrow 0$ when $a\rightarrow 0$. The same estimate is valid also for
diagrams including the propagators (\ref{d9}) - (\ref{d11}) provided $M_r
< \lambda N^\gamma$.

In the domain $V_{out}$ the corresponding expression for $\Pi^L_{out}$
qafter rescaling the integration variables $ap=u$ may be written in the form

\begin{equation}
\Pi_{out}^L \sim N^{L+4} \int_{V_{out}} \, du \Tr \{ \Bigl(
1-\gamma_5\Bigr) \biggl[ \frac {(1+\Gamma_{11}) }{2}  V^{i_L} S(u+
aQ_{L-1}) \ldots V^{i_1} S(u) \biggr] \label{d16} \end{equation} $$
 +\sum_{r, \pm} \frac {(1\pm \Gamma_{11}) }{2} V^{i_L} S_r (u+ aQ_{L-1})
\ldots V^{i_1} S_r (u) \}
$$

Let us consider the first term in eq.(\ref{d16}).

Near the edges of Brilluen zones $u=(\pi ,0,0,0)$ e.t.c. the vertex
function $V^i\sim O(1)$ whereas the propagators may be written as follows

\begin{equation}
S(u) \sim a \biggl( \frac {\hat{u}}{u^2+ \mu^2} \biggr)
\label{d17}
\end{equation}
where $\mu^2 \sim O(1)$. One sees that the diagrams with arbitrary number
of external Higgs lines are divergent in the limit $a\rightarrow 0$,
leading to nonrenormalizability of the theory. Higher derivative terms
also arise. In particular $\Pi^2$ contains a finite term $\sim k^4$. It
illustrates a nondecoupling of heavy particles in the model with the
Yukawa coupling of the order of the cut-off.

However the presence of the PV fields changes the situation drastically.
The PV fields produce the analogous contributions to eq.(\ref{d16}) which
compensate the contribution of the original fields.

For definiteness we consider now the case of even number of generations,
e.g. $l=2$. As we have already mentioned the Higgs meson scattering
amplitudes eq.(\ref{d16}) contain the contribution of the positive chirality
physical fields $\psi^l_+$, and PV fields $\psi_r ,\chi_r$ of both
chiralities. Now we shall show that the contributions of the positive and
negative chirality fields are equal in the limit $a\rightarrow 0$.

Indeed the trace over the $SO(10)$ indices in the eq.(\ref{d15}) is
proportional to

\begin{equation}
\Tr [(1\pm \Gamma_{11} )\Gamma_{i_1} \ldots \Gamma_{i_L} ]
\label{d18}
\end{equation}
\begin{equation}
\Gamma_{11} \sim \varepsilon^{l_1\ldots l_{10}} \Gamma_{l_1} \ldots
\Gamma_{l_{10}}
\label{d19}
\end{equation}
where $\varepsilon^{l_1\ldots l_{10}}$ is antisymmetric with respect to the
permutations of any two indices. The term proportional to $\Gamma_{11}$ is
different from zero only if the product of $\Gamma_i$ can be reduced to
the sum of the terms proportional to the product $\Gamma_1 \ldots
\Gamma_{10}$.

In the limit $a\rightarrow 0$ the contribution of this term to the
scattering amplitudes to the leading order is proportional to
\begin{equation}
a^{-4}\int\, dk_1\ldots dk_{10} \varphi^{i_1} (k_1) \ldots \varphi^{i_{10}}
(k_{10})  \varepsilon^{i_1 i_2 \ldots i_{10}} \delta (k_1 +\ldots k_{10})
\label{d20}
\end{equation}
This term is zero due to antisymmetry of $\varepsilon^{i_1\ldots i_{10}}$.
Next to the leading term is proportional to $a^{-2}$ and contains two
additional momenta $k_{i_l},k_{i_m}$, and there is also a finite term with
four momenta $k_{i_l}k_{i_m}k_{i_r}k_{i_s}$. All other terms vanish in the
limit $a\rightarrow 0$. The terms with two and four momenta are zero due
to antisymmetry of $\varepsilon^{i_1\ldots i_{10}}$.

Therefore one can drop the terms proportional to $\Gamma_{11}$  in the
eq.(\ref{d16}) and instead of summing over $\pm$ to multiply the last term
by the factor 2. Now the first and the second terms in the eq.(\ref{d16})
have the same structure, and expanding this equation in terms of $M^2_r$
one gets
   \begin{equation}
\Pi_L \sim 2 \{ \frac 1{a^4} \bigl( \sum_r C_r + 1\bigr) f_1(k) +
\frac 1{a^2} \bigl( \sum_r C_r M_r^2 \bigr) f_2(k)
\label{d21}
\end{equation}

$$
+\bigl( \sum_r C_r M_r^4 \bigr) f_3(k) \} + O(a)
$$
Assuming the PV conditions
\begin{equation}
\sum C_r +1=0; \qquad \sum C_r M_r^2 =0; \qquad \sum C_r M_r^4 =0;
\label{d22}
\end{equation}
we see that $\Pi^L \rightarrow 0$ when $a\rightarrow 0$.

No new problems arise for the mixed diagrams including scalar and vector
external lines. Their contribution also vanish in the continuum
limit (in fact it follows from the gauge invariance of the theory and
vanishing of the diagrams with only scalar external lines).

Therefore all spinorial loops generated by the Yukawa--Wilson interaction
do not contribute in the continuum limit and the heavy particles decouple.

In the case of an odd number of generations the same conclusion holds if
one introduces the infinite number of PV fields following the procedure
described in \cite{SF1}, \cite{SF}. One should take in the
eq.s(\ref{d2},\ref{d3}) the summation over $r$ from  $-\infty$  to
$+\infty$ ($r>0$ corresponding to the positive chirality PV fields,
$r<0$ to the negative chirality PV fields and $r=0$ to the original field
$\psi^+$) and put $M_r=M|r|$. The analytic expression for the spinor
loop with $L$ external lines can be written in the same way as in
ref.\cite{SF} and after summation over $r$ looks as follows
\begin{equation} \Pi_L\sim \int_{-\frac {\pi}{a}}^{\frac {\pi}{a}}
\sum_{l=0}^{n-1} \frac {A_l(p, Q_l)}{\sqrt{M^2 \left( s^2 +m^2\right)
\sinh \left( \frac {\pi \sqrt{s^2+m^2}}M \right) }} d^4p \label{d23}
\end{equation}
where $Q_l=k_1+\ldots +k_l$, $k_l$ are the external momenta, and $A_l$ are
some polinomials over $p,Q_l$. This expression decreases exponentially
when $a\rightarrow 0$ providing decoupling of heavy particles in the odd
number generation case as well. (As was indicated in the ref.\cite{SF} in
the lattice model it is not necessary to take an infinite number of the PV
fields. It is sufficient to take $|r|\leq  \tilde{N}(a)$ where
$\tilde{N}(a)\rightarrow \infty$ when $a\rightarrow 0$.

Up to now we considered only spinorial loops. However the diagrams with
internal lines of Higgs and Yang-Mills fields generated by the
Yukawa-Wilson coupling are also proprtional to the large parameter $N$ and
for these diagrams the compensation mechanism described above does not
work. If one tries to develop a perturbation expansion for the action
(\ref{d2}) one finds an infinite series of divergent diagrams indicating
the failure of weak coupling expansion. This decease may be cured by the
method analogous to the one used in ref\cite{SF}. The Yang-Mills and Higgs
actions should be modified by introducing higher order covariant
derivatives of sufficiently high order.By choosing appropriately the
parameter $\Lambda$ multiplying the higher derivative term in the modified
action one can make the contribution of these diagrams vanishing in the
continuum limit.

This mechanism may be illustrated by the analysis of the fermion
selfenergy diagram generated by the Yukawa-Wilson coupling. The
corresponding integral looks as follows
\begin{equation}
\Sigma \sim N^2 \int_{-\frac {\pi}{a}}^{\frac {\pi}{a}}
\Bigl(1-\gamma_5\Bigr) \Bigl(1+\Gamma_{11}\Bigr)V^iS(p+Q)D(p)V^k
d^4p \label{d24}
\end{equation}
Here $D(p)$ is a modified Higgs field propagator corresponding to
the sixth order derivatives in the regularised action
\begin{equation}
D = \Bigl[a^{-2} \sum_{\mu} \Bigl( \cos (p_\mu a)-1 \Bigr) + \Lambda^2
\Bigl(\sum_{\mu}( \cos (p_\mu a)-1) \Bigr)^3 \Bigr] \label{d25}
\end{equation}
$S(p)$ is the fermion propagator (\ref{d8}) and $V^i$ is the interaction
vertex (\ref{d6}).

To analyse this integral we separate as before the integration domain
into $V_{in}$ and $V_{out}$ according to the eq.(\ref{d14}). In the
domain $V_{in}$ one can use the expansion over $(pa)$ and the integral
looks as follows (omitting irrelevant matrix factors)
\begin{equation}
\Sigma_{in} \sim N^2 \int_{V_{in}} \frac {(q+p)^4a^4(q+p)}{ \Bigl[(q+p)^2 +
(v\lambda^{-1})^2a^2(q+p)^4 \Bigr] \Bigl[p^2 + \Lambda^2a^6p^6
\Bigr]} \label{d26}
\end{equation}
In the limit $a \rightarrow 0$, $\Sigma_{in}< \lambda N^{2-5\gamma}
\rightarrow 0$.  In the domain $V_{out}$ after rescaling the integration
variables $ap_\mu = u_\mu$ one has \begin{equation} \Sigma_{out} \sim
a^{-2}N^2 \int_{V_{out}} \frac {s(u+Qa)V(u+Qa)V(u+Qa)}{ \Bigl[s^2(u+Qa) +
m^2(u+Qa) \Bigr] \Bigl[( \cos(u) - 1) + \Lambda^2a^2( \cos(u) - 1)^3
 \Bigr]} \label{d27} \end{equation} For small $a$, $\Sigma_{out} <
 \lambda^3N^{-5}\Lambda^{-2}$. Choosing $\Lambda^2 = \lambda^2N^{\beta}, 5
  \ll \beta \ll 6$ we can make this term vanishing in the
 continuum limit.  Exactly the same arguments are applied to the
 selfenergy diagram with the fermion propagator (\ref{d9}) and all other
 diagrams with internal Higgs or Yang-Mills lines. We note that the
 condition $\Lambda^2 \ll \lambda^2N^5$ means that the masses
 $M_{\Lambda}$ of additional exitations which appear due to the presence
 of higher derivative regulators are less than $\lambda N^{ \frac
 {1}{4}}$. In particular one can take $M_\Lambda \sim M_r$ where
 $M_r$ are the PV masses. Both $M_\Lambda$ and $M_r$ correspond to
 the poles in the domain $V_{in}$.

 \section {Discussion}

The construction discussed above is obviously applicable also to the
standard model as the gauge group of the standard model is a subgroup of
$SO(10)$, and to get its lattice version it is sufficient to put in our
action all vector fields except for the gluons and electroweak bosons
equal to zero.

In the scaling limit our model coincides essentially with the model of ref
\cite{SF} if one introduces to the latter a gauge invariant interaction of
Higgs fields. However it has an advantage of being manifestly gauge
invariant for a finite lattice spacing, which makes it more suitable for
nonperturbative calculations. In particular there is no problem of
nonperturbative gauge fixing.

The gauge invariant action we propose is given by
the eq.  (\ref{d2}) written in terms of shifted Higgs fields $\varphi
\rightarrow \varphi + v$ and modified by introducing higher derivative
kinetic term for for the Higgs and Yang-Mills fields We proved that the
weak coupling expansion near the ground state corresponding to the
quadratic piece of this action reproduces in the continuum limit a
manifestly gauge invariant perturbation theory for the $SO(10)$ model. As
was discussed above the perturbation theory for the standard model is
obtained by reducing the $SO(10)$ group to the subgroup $SU(3) \otimes
SU(2) \otimes U(1)$.

In this paper we made no attempts to study a real phase structure of the
lattice model which requires esentially nonperturbative treatment.It
remains to be seen if our conclusions are valid beyond the perturbation
theory. We note that there is an important difference between our
construction and the original Smit-Swift model. In the original model the
physical reason for the absence of interacting chiral fermions was the
formation  of scalar-fermion bound states due to the strong Yukawa-Wilson
coupling. In our case, as has been shown in the previous section, the
effective Yukawa-Wilson interaction is weak (vanishing in the scaling
limit). The effective interaction between Higgs mesons induced by the
fermion loops is supressed by the PV fields, and the interaction induced
by the exchange of Higgs fields is supressed by the higher derivative
regulators. In fact the only relevant domain in the momentum space in our
model is the domain near $p=0$ ($V_{in}$ as defined by the eq.
\ref{d14}). In this domain the doublers are absent and the Yukawa-Wilson
interaction is negligible.

We repeat once again that the most important problem is to develop for the
analysis of this model a reliable nonperturbative approach which allows
to study the phase structure of the theory and to check if the conclusions
about the doublers decoupling remain valid and if the light fermions
interact nontrivially in the continuum limit.Our analysis has been done in the
framework of the weak coupling expansion. We showed that although the
Yukawa--Wilson coupling in our model is large, perturbation theory can be
applied. In comparison with the model of ref.\cite{SF} the present
construction has an advantage of being manifestly gauge invariant for a
finite lattice spacing which makes it more suitable for nonperturbative
calculations. In particular there is no problem of nonperturbative gauge
fixing. However the most important problem is to develop for the analysis
of this model a reliable nonperturbative approach and to check if the
conclusion about the doublers decoupling remains valid outside of
perturbation theory and if the light fermions interact nontrivially in the
continuum limit.

$$~$$ {\bf ACKNOWLEDGMENT} $$~$$ I am indebted to G.A.Kravtzova for the
help in preparing the manuscript.
  $$~$$  
\begin{thebibliography}{99} {\small
\bibitem{SF} S.A.Frolov, A.A.Slavnov, {\it Removing fermion doublers in
chiral gauge theories on the lattice}, Max-Planck Inst. Preprint
MPI--Ph 93-12; hep-lat/9303004.  \bibitem{Kars} L.H.Karsten, in {\it Field
theoretical methods in particle physics. (ed. by W.Ruhl) Plenum Press, New
York (1980).}  \bibitem{Smit1} J.Smit, Nucl.Phys.B175 (1980) 307.
\bibitem{Swift} P.D.V.Swift,Phys.Lett.B145 (1984) 256.  \bibitem{Smit2}
J.Smit, Acta Phys.Polonica B17 (1986) 531.  \bibitem{Smit3}J.Smit,
Nucl.Phys.B (Proc. Suppl.)17 (1990) 3.  \bibitem{Golt} M.F.L.Golterman,
Nucl.Phys.B (Proc. Suppl.) 20 (1991) 528.  \bibitem{Pet} D.N.Petcher, {\it
Chiral gauge theories and fermion-Higgs systems} Preprint of Washington
University, 1993.  \bibitem {GPS} M.F.L.Golterman, D.N.Petcher, J.Smit,
Nucl.Phys.B370 (1992) 51.  \bibitem{BS} W.Bock, A.K.De, J.Smit,
Nucl.Phys. B388 (1992) 243.  \bibitem{GP} M.F.L.Golterman, D.N.Petcher,
Nucl.Phys.B (Proc.Suppl.) 26 (1992) 483.  \bibitem{SF1} {\it An invariant
regularization of the standard model} Saclay Preprint SPhT/92-051;
Pys.Lett.B309 (1993) 344.} \end{thebibliography}
\end{document}